\documentclass[fleqn,twoside]{article}
\usepackage[headings]{espcrc2}

\readRCS
$Id: espcrc2.tex,v 1.2 2004/02/24 11:22:11 spepping Exp $
\usepackage{graphicx}
\usepackage[figuresright]{rotating}

\newcommand{\AmS}{{\protect\the\textfont2
  A\kern-.1667em\lower.5ex\hbox{M}\kern-.125emS}}

\title{Rotational energy transfer in H$_2$+H$_2$}

\author{Renat A. Sultanov\address[BCRL]{Business Computing Research Laboratory, St. Cloud State University,
31 Centennial Hall, 720 Fourth Avenue South, St Cloud, MN 56301-4498}
\thanks{sultanov@bcrl.stcloudstate.edu; r.sultanov2@yahoo.com},
Dennis Guster\addressmark\thanks{dcguster@stcloudstate.edu}.}

\runtitle{Rotational energy transfer}
\runauthor{R.A. Sultanov and D. Guster}

\begin{document}

\begin{abstract}
Quantum-mechanical close-coupling calculations for state-to-state cross sections and thermal rates are reported 
for H$_2$+H$_2$ collisions. Two recently developed
potential energy surfaces (PES) for the H$_2-$H$_2$ system are applied, namely,
the global potential surface from the work of A.I. Boothroyd, P.G. Martin, W.J. Keogh, M.J. Peterson,
J. Chem. Phys., 
116 (2002) 666, and a restricted, model surface from the works of P. Diep, J.K. Johnson, J. Chem. Phys., 
113 (2000) 3480; ibid. 112, 4465. The low temperature limit is investigated. We found significant differences
in cross sections and corresponding thermal rates calculated with these two PESs.
\end{abstract}

\maketitle

\section{Introduction}

The investigation of elastic and inelastic collisions between molecules and between molecules and
atoms can provide valuable information about interactions, chemical properties and energy transfer dynamics
[1-15].
The hydrogen molecule is the simplest and most abundant molecule in the universe's   
molecular clouds and plays an important role in many areas of astrophysics.
For example, knowledge of the ro-vibrational excitation and de-excitation rate constants
in molecular hydrogen collisions is of fundamental importance for understanding and modeling the energy 
balance in the interstellar medium. 
The energy transfer processes involving H$_2$ molecules control  
the evolution of shock fronts 
and photodissociation regions (PDRs) in the interstellar medium.
%
Additionally, the energy transfer between H$_2$ molecules and between H$_2$ and other atoms/molecules 
is important for 
cooling of primordial gas and shock wave-induced heating in the interstellar media.
However,
to accurately model the thermal balance and kinetics of such important systems one needs accurate
state-to-state rate constants $k_{vjv'j'}(T)$.


Experimental measurement of quantum state resolved cross sections and rates is a very difficult 
technical problem. 
On the other hand, accurate theoretical data requires precise potential energy surfaces
and reliable dynamical treatment of the collision processes.
%
The first attempt to construct a realistic full-dimensional ab initio PES for the H$_2-$H$_2$
system was done in works \cite{schwenke88},
and the potential was widely used in a variaty of methods  and computation techniques.

Recently the H$_2$$-$H$_2$ system has been reinvestigated and
an accurate interaction potential from the first principles has been developed in work \cite{diep00}. 
However, in this work the Diep and Johnson potential energy surface (DJ PES) was extrapolated only for the rigid
rotor monomer model of H$_2$$-$H$_2$.

An extensive study of the H$_2$$-$H$_2$ PES has been recently reported by Boothroyd et al.
\cite{booth02}, where the potential energies have been represented at 48180 geometries respectively
with a large basis set at the multireference configuration interaction level.

In this work we provide a test of sensitivity for the new PESs
for collisions of rotationally excited H$_2$ molecules
\begin{equation}
\mbox{H}_2(j_1) +\mbox{H}_2(j_2) \rightarrow \mbox{H}_2(j'_1) + \mbox{H}_2(j'_2).
\label{eq:h2h2}
\end{equation}
We apply only the new surfaces mentioned above.

The scattering cross sections and their
corresponding rate coefficients are calculated using a non reactive quantum-mechanical close-coupling
approach. In the next section we will briefly outline the method. Our results and discussion are presented in 
Section 3. 

\section{Method}

In this section we provide a brief outline of the quantum-mechanical close-coupling approach used 
in our calculations.
All relevant details have been developed in works \cite{green75}.
The hydrogen molecules are treated as linear rigid 
rotors. The model has been applied in few previous works, see for example \cite{green75,flower98a}. 
For the considered range of kinetic energies of astrophysical interest 
the rotor model is considered to be adequate \cite{bourlot99}.

The cross sections for rotational excitation and relaxation phenomena can be obtained directly 
from the $S$-matrix.
In particular the cross sections for excitation from $j_1j_2\rightarrow j'_1j'_2$ summed over 
the final $m'_1m'_2$
and averaged over the initial $m_1m_2$ corresponding projections of the H$_2$ molecules angular 
momenta $j_1$ and $j_2$ 
are given by
\begin{eqnarray}                               
\sigma(j'_1,j'_2;j_1j_2,\epsilon)=\frac{\pi}{(2j_1+1)(2j_2+1)k_{\alpha\alpha'}}\nonumber \\
\sum_{Jj_{12}j'_{12}LL'}(2J+1)|\delta_{\alpha\alpha'}-
S^J_{\alpha \alpha'}(E)|^2.
\label{eq:cross}
\end{eqnarray}
The kinetic energy is 
%
$\epsilon=E-B_1j_1(j_1+1)-B_2j_2(j_2+1)$.
%
Here $E$ is the total energy in the system,
$B_{1(2)}=60.8\hspace{2mm}\mbox{cm}^{-1}$
are the rotation constants of the colliding H$_2$ molecules,
$J$ is total angular momenta of the
4-atomic system, $\alpha \equiv (j_1j_2j_{12}L)$, where $j_1+j_2=j_{12}$ and $j_{12}+L=J$,
$k_{\alpha \alpha'}=2M_{12}(E+E_{\alpha}-E_{\alpha'})^{1/2}$ is the channel wavenumber and
$E_{\alpha(\alpha')}$ are rotational channel energies.

The log-derivative matrix is propagated to large $R$-intermolecular distances, since all experimentally observable
quantum information about the collision is contained in the asymptotic behaviour of functions 
$U^{JM}_{\alpha}(R\rightarrow\infty)$. 
A set of coupled second order differential equations
for the unknown radial functions $U^{JM}_{\alpha}(R)$ is solved
\begin{eqnarray}
\left(\frac{d^2}{dR^2}-\frac{L(L+1)}{R^2}+k_{\alpha}^2\right)U_{\alpha}^{JM}(R)=2M_{12}\nonumber \\
\sum_{\alpha'}
\int <\phi^{JM}_{\alpha}(\hat r_1,\hat r_2,\vec R)
|V(\vec r_1,\vec r_2,\vec R)|\nonumber \\
\phi^{JM}_{\alpha'}(\hat r_1,
\hat r_2,\vec R)>U_{\alpha'}^{JM}(R) d\hat r_1 d\hat r_2 d\hat R,
\label{eq:cpld}
\end{eqnarray}
where $\alpha \equiv (j_1j_2j_{12}L)$.
We apply the hybrid modified log-derivative-Airy propagator in the general purpose scattering program MOLSCAT 
\cite{hutson94} to solve the coupled radial equations (\ref{eq:cpld}). Different propagator schemes included 
in MOLSCAT have been tested. Our calculations showed that other propagators can also provide quite stable results.

The numerical results are matched to the known asymptotic solution to 
derive the physical scattering $S$-matrix
%
%
\begin{eqnarray}                              
U_{\alpha}^J
\mathop{\mbox{\large$\sim$}}\limits_{R \rightarrow + \infty}
\delta_{\alpha \alpha'}
e^{-i(k_{\alpha \alpha}R-(l\pi/2))}- \nonumber \\
\left(\frac{k_{\alpha \alpha}}{k_{\alpha \alpha'}}\right)^{1/2}
S^J_{\alpha \alpha'} e^{-i(k_{\alpha \alpha'}R-(l'\pi/2))},
\end{eqnarray}
%
The method was used for each partial wave until a converged cross section was obtained. 
It was verified that the results are converged with respect to the number of partial waves as well as
the matching radius, $R_{max}$, for all channels included in our calculations.

The relationship between the rate coefficient $k_{j_1j_2\rightarrow j'_1j'_2}(T)$ and the 
corresponding
cross section $\sigma_{j_1j_2\rightarrow j'_1j'_2}(E_{kin})$ can be obtained through the following
weighted average
\begin{eqnarray}
k_{j_1j_2\rightarrow j'_1j'_2}(T) = \frac{8k_BT}{\pi\mu}\frac{1}{(k_BT)^2}
\int_{\epsilon_s}^{\infty}\epsilon d\epsilon\nonumber \\
\sigma_{j_1j_2\rightarrow j'_1j'_2}(\epsilon)e^{-\epsilon/k_BT},
\end{eqnarray}
where $\epsilon = E - E_{j_1} - E_{j_2}$ is precollisional translational energy at the
translational temperature $T$, $k_B$ is Boltzman constant, $\mu$ is reduced mass of the 
molecule-molecule system and $\epsilon_s$ is the minimum kinetic energy for the levels $j_1$ and $j_2$
to become accessible.

\section{Results}

As we mentioned in the Introduction, in this work
we apply the new PESs from works \cite{diep00} and \cite{booth02}.
The DJ PES \cite{diep00} is constructed for
the vibrationally averaged rigid monomer model of the H$_2$$-$H$_2$ system to the complete basis set limit using 
coupled-cluster theory with single, double and triple excitations. A four term spherical harmonics expansion 
model was chosen to fit the surface. It was demonstrated, that
the calculated PES can reproduce the quadrupole moment to within 0.58 \% and 
the experimental well depth to within 1 \%.

The bond length was fixed at 1.449 a.u. or 0.7668 \r{A}. DJ PES is 
defined by the center-of-mass intermolecular distance, $R$, and three angles: $\theta_1$ and $\theta_2$ are the 
plane angles and $\phi_{12}$ is the relative torsional angle. The angular increment for each of the three angles 
defining the relative orientation of the dimers was chosen to be $30^{\circ}$. 

The BMKP PES \cite{booth02} is a global six-dimensional potential energy surface for two hydrogen molecules.
It was especially constructed to represent the whole interaction region of the chemical reaction dynamics of the 
four-atomic system and to provide an accurate as possible van der Waals well.
%
In the six-dimensional conformation space of the four atomic system the conical intersection forms a complicated 
three-dimensional hypersurface. 
%
Because the BMKP PES uses cartesian coordinates to compute distances between 
four atoms, we have devised some fortran code, which converts spherical coordinates used in
Sec. 2 to the corresponding cartesian coordinates and computes the distances between the four atoms. In all our
calculations with this potential the bond length was fixed at 1.449 a.u. or 0.7668 \r{A} as in the DJ PES.

A large number of test calculations have also been done to secure the convergence of the results with 
respect to all 
parameters that enter into the propagation of the Schr\"odinger equation \cite{renat06}. 
This includes the intermolecular distance $R$, the total angular momentum $J$ of the four atomic system, $N_{lvl}$ 
the number of rotational levels to be included in the close coupling expansion and others
(see the MOLSCAT manual \cite{hutson94}).
We reached convergence for the integral cross sections, $\sigma(E_{kin})$, in all considered collisions. In the 
case of the
DJ PES the propagation has been done from 2 \r{A} to 10 \r{A}, since this potential is defined only for 
those specific distances. For the BMKP PES we used $r_{min}=1$ \r{A} to $r_{max}=30$ \r{A}. We also applied a few 
different propagators included in the MOLSCAT program.


Now we present our results for different
rotational transitions in collisions between $para/para$- and ortho-/ortho-hydrogen molecules (\ref{eq:h2h2}).
%
Our main goal in this work is first: to carry out complete quantum-mechanical calculations with new potentials
for different transitions
in $p$-H$_2$+$p$-H$_2$ and $o$-H$_2$+$o$-H$_2$ collisions and, second, to provide a comparative study and check
the sensitivity of the two newest surfaces for the very important and fundamental H$_2$-H$_2$ system.

The energy dependence of the elastic integral cross sections on the total energy $\sigma_{el} (E_{tot})$
in the H$_2$+H$_2$ system is represented in Fig.\ 1 (upper plots) together with
the state-resolved integral cross sections $\sigma_{j_1j_2\rightarrow j'_1j'_2}(E_{tot})$
for the $j_1=j_2=0 \rightarrow j'_1=2,j'_2=2$ and $j_1=j_2=1 \rightarrow j'_1=1,j'_2=3$ rotational transitions
(lower plots) for both the BMKP and the DJ PESs respectively. As can be seen both PESs provide the
same type of the behaviour in the cross section. 
These cross sections
are in basic agreement with recent time-dependent quantum-mechanical calculations in work
\cite{guo02}. However, our results show, that the
DJ PES generates about 30 \% higher values for the cross sections relatively to the BMKP PES.

\begin{figure}
\begin{center}
\includegraphics*[width=18pc]{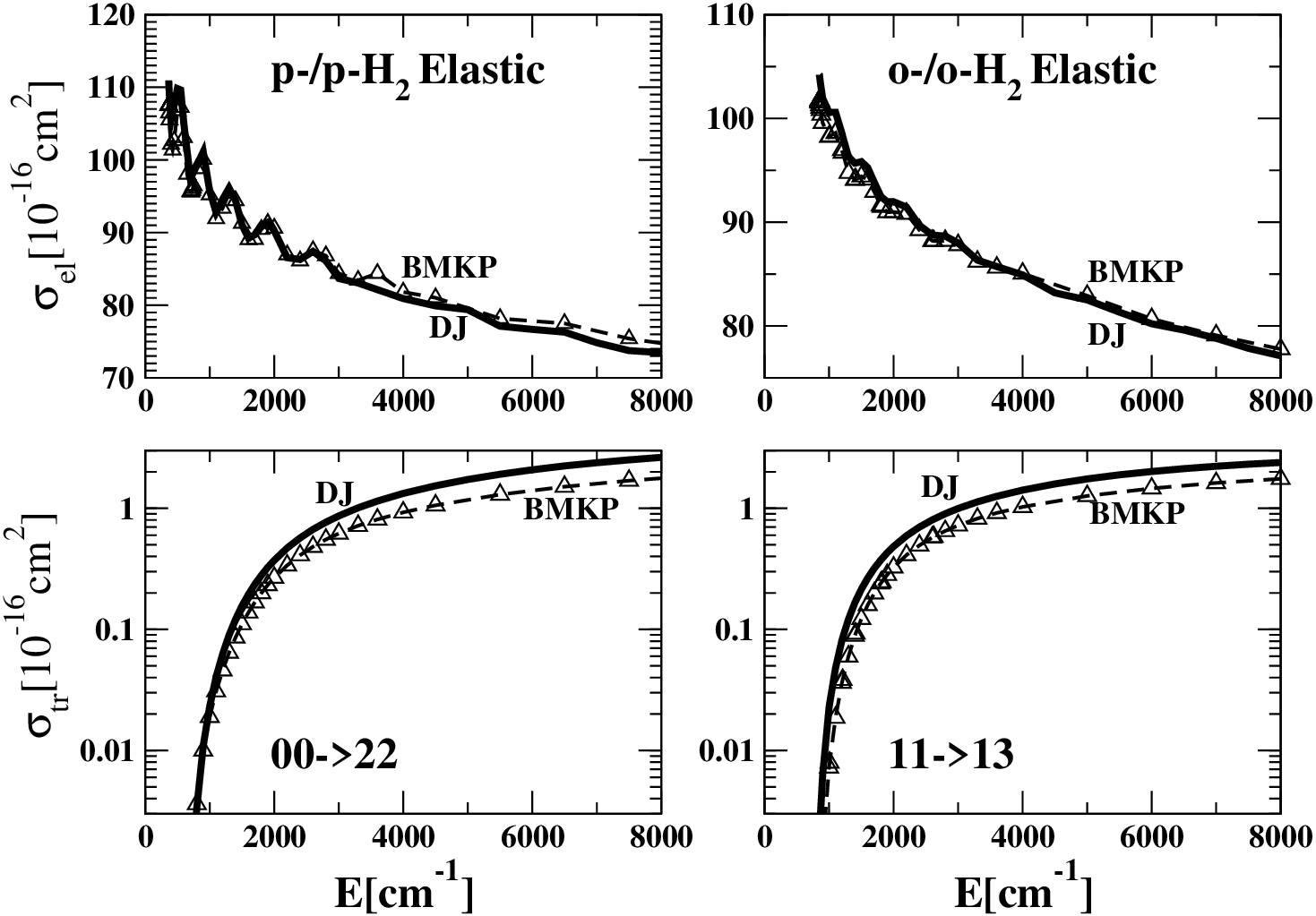}
\end{center}
\caption{Rotational state resolved integral cross sections for elastic scattering in the case of para-/para- and
ortho-/ortho-hydrogen and transitions, when $j_1=j_2=0 \rightarrow j'_1=2,j'_2=2$ and
$j_1=j_2=1 \rightarrow j'_1=1,j'_2=3$. Calculations are done with the DJ (bold lines)
and BMKP (open triangles up) PESs. The compensating factor 
of 2 is included only in the elastic cross sections.}
\label{fig:fig1}
\end{figure}       

Also, it is important to point out
here, that for comparison purposes we don't include the compensating factor of 2 mentioned in \cite{flower87}.
However, in Fig.\ 2 (left plot) and in our subsequent calculations of the thermal rate coefficients, 
$k_{jj'}(T)$, the factor is included.

Significant differences in the cross sections of the two potentials are reflected in the state-resolved
transition states $j_1=0,j_2=0 \rightarrow j'_1=2,j'_2=0$, as shown in Fig.\ 2 (right panel). 
That is why
it seems that the DJ PES can provide much better results, as seen in the same figure in the left panel.
Specifically, when we
present the results for the corresponding thermal rates $k_{00-20}(T)$ calculated with the DJ potential 
together with results of other theoretical calculations \cite{flower98a} and \cite{flower98b} the agreement 
is almost perfect.
Next, Fig.\ 3 provides the same results for the cross sections,
but includes the details at low energies,
together with the corresponding experimental data from \cite{mate05}.
Now one can better see the considerable differences in these two cross sections.
However, the DJ PES is able to provide
very good agreement with experimental data \cite{mate05}.
Thus we conclude, that DJ PES is much better suited for the H$_2-$H$_2$ system.
Moreover, in Fig.\ 4 we provide thermal rates for different transition
states calculated with only the DJ PES and in comparison with other theoretical 
data obtained within different dynamical methods and PESs. 
Again the agreement is very good.


\begin{figure}
\begin{center}
\includegraphics*[width=18pc]{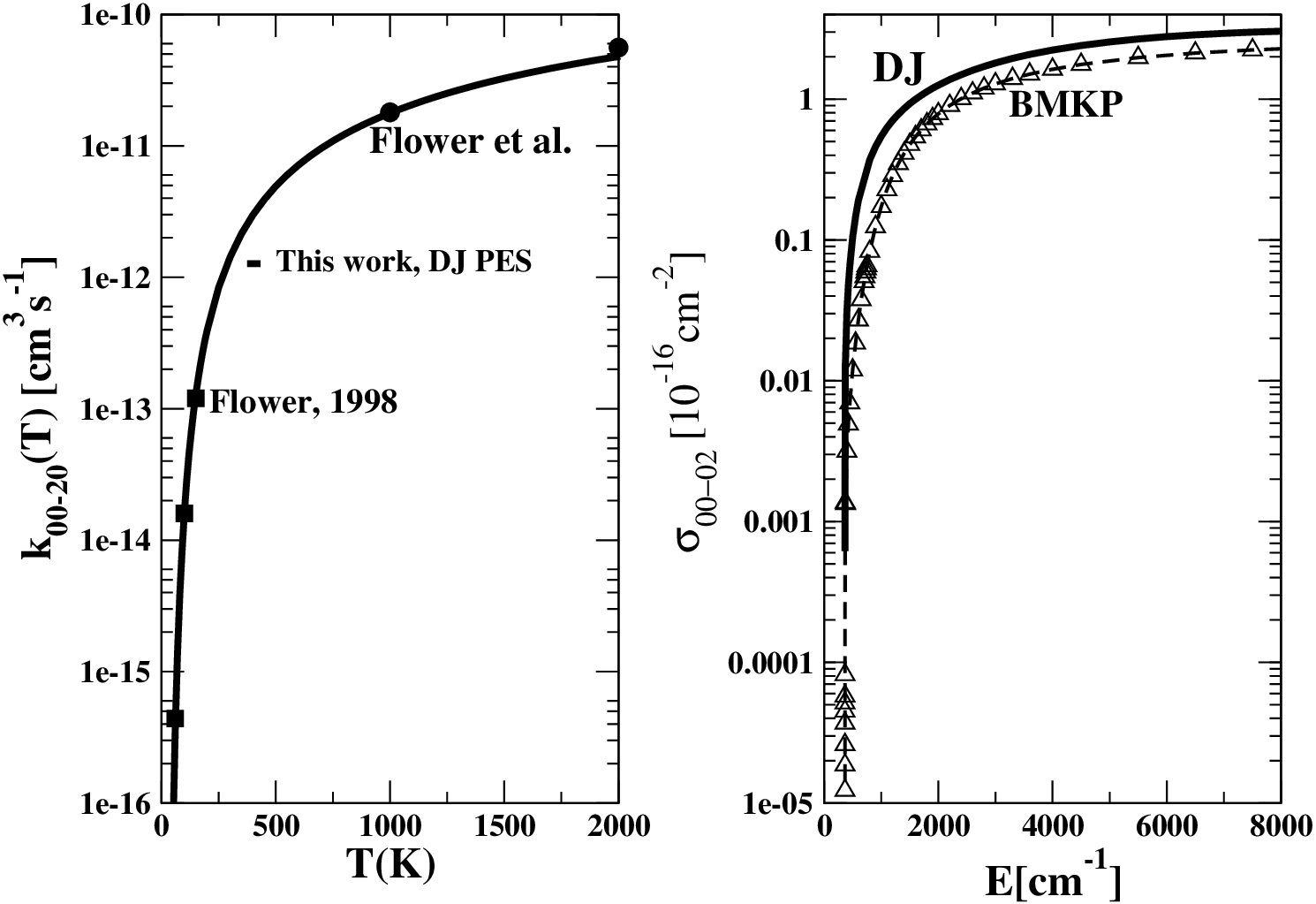}
\end{center}
\caption{Temperature dependence of the state-resolved thermal rate constant (left panel) and corresponding 
cross sections (right panel) for the transition $j_1=j_2=0 \rightarrow j'_1=2,j'_2=0$. 
%
%
Squars and circles are the data from \cite{flower98a} and \cite{flower98b} respectively. 
Our results with the DJ PES are depicted with
solid lines, the open triangles up represent our results with the BMKP surface.}
\label{fig:fig2}
\end{figure}

\begin{figure}
\begin{center}
\includegraphics*[width=18pc]{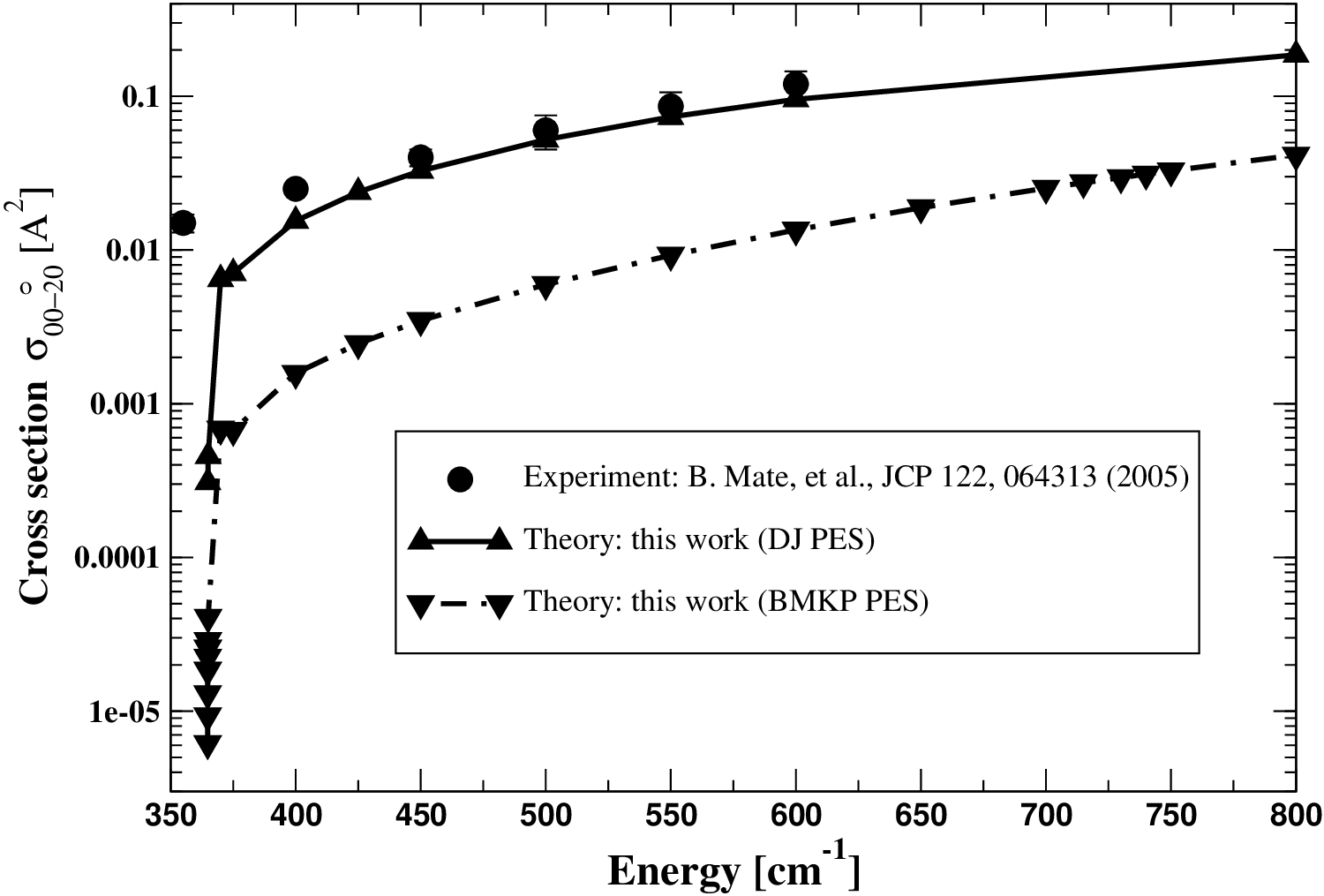}
\end{center}
\caption{Cross sections for the 00$\rightarrow$20 rotational transition calculated with the DJ and BMKP PESs 
for the H$_2$+H$_2$ collision. The circles are some experimental data from the
work \cite{mate05}, triangles up and 
triangles down are the results of this work using the DJ and BMKP PESs respectively.}
\label{fig:fig3}
\end{figure}

Finaly,
Table 1 depicts the thermal rates $k_{00\rightarrow 20}(T)$ and $k_{02\rightarrow 22}(T)$
at lower kinetic temperatures, up to 10 K. Numbers in each column should be multiplied by
corresponding prefactor $P$.
As can be seen our results are in very good agreement with recent experimental and theoretical data at higher
temperatures. However, at $T<30$ K
we could not reproduce effectively the results of \cite{mate05}. It may be because of
the interatomic (H$_2$-H$_2$) distance
restrictions in the DJ surface in which the potential ranges from 2 \r{A} to only 10 \r{A}. 
It seems plausible, that
for such low energies one needs to carry out calculations even beyond the 10 \r{A} limit.


We provide close-coupling quantum-mechanical calculations of the state-resolved rotational 
excitation and deexcitation cross sections and rates
for molecular hydrogen collisions. 
A test of convergence and the results for cross sections and rate coefficients
using two different potential energy surfaces for the H$_2-$H$_2$
system have been obtained for a wide range of kinetic energies \cite{renat06}.
%


Our calculations revealed, that both PESs can provide the same type of 
behaviour in regard to cross sections  and rates for different transition states.
However, significant differences in the cross section of the 00$ \rightarrow$ 20 transition have been found.
Also, it was already indicated, that at even larger kinetic energies the DJ potential overestimates relative 
to the BMKP surface the cross sections and rates by about 20-40 \% \cite{renat06}.
%
%
%

In conclusion, the results of these calculations show, that additional
work is needed to further improve the BMKP PES, particularly the part of the surface, which is responsible
for the 00$ \rightarrow$ 20 transition. We also found, as can be verified
from Table 1,
it is difficult to use the DJ surface and carry out reliable calculations at temperatures lower than 30 K.



\begin{table}[htb]
\caption{Thermal rates $k_{00\rightarrow 20}(T)$ and
$k_{02\rightarrow 22}(T)$ (m$^3$s$^{-1}$) at low and very low temperatures calculated with the DJ surface
in comparison with available experimental$^{*)}$ and theoretical data from \cite{mate05}}
%
\label{table:1}
\begin{tabular}{lcccccc}
\hline
&\multicolumn{3}{c}{$k_{00\rightarrow 20}$} & \multicolumn{2}{c}{$k_{02\rightarrow 22}$} \\
\hline
$T (\mbox{K})$ &    DJ   & \cite{mate05}$^*$ &  \cite{mate05}   &    DJ    & \cite{mate05}    &   $P$        \\
\hline
100            &  1.64   &  2.2(4)           &  1.92            & 2.63     & 3.23             &   10$^{-20}$ \\
60             &  3.81   &  6.0(7)           &  4.75            & 5.62     & 7.97             &   10$^{-22}$ \\
30             &  0.54   &  1.1(1)           &  0.80            & 0.65     & 1.34             &   10$^{-25}$ \\
20             &  0.87   &  2.3(3)           &  1.63            & 0.93     & 2.71             &   10$^{-29}$ \\
10             &  0.38   &  2.7(7)           &  1.65            & 0.33     & 2.62             &   10$^{-40}$ \\
%
\hline
\end{tabular}\\[2pt]
\end{table}

\begin{figure}
\begin{center}
\includegraphics*[width=18pc]{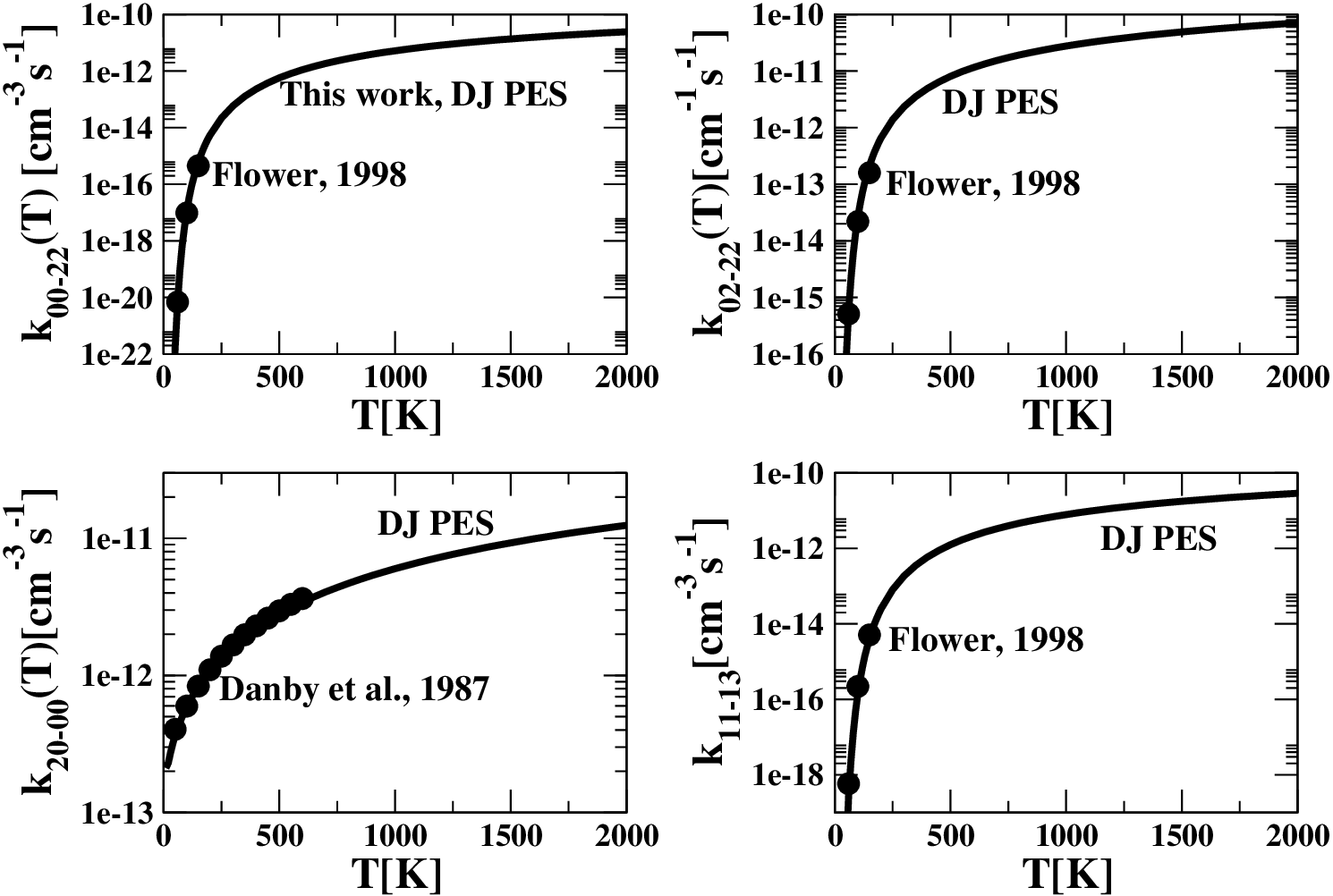}
\end{center}
\caption{  
Thermal rates
calculated with the DJ PES for the $00 \rightarrow 22$, $02 \rightarrow 22$, 
$20 \rightarrow 00$ and $11 \rightarrow 13$ transitions and other results
\cite{flower87,flower98a}.}  
\label{fig:fig4}
\end{figure}       

\end{document}